\documentclass[nofootinbib,twocolumn,showpacs,preprintnumbers,amsmath,amssymb,prc]{revtex4}

\usepackage{graphicx}% Include figure files
\usepackage{dcolumn}% Align table columns on decimal point
\usepackage{bm}% bold math

\begin{document}

\title{Nuclear mass number dependence of inclusive production of $\omega$ and $\phi$ mesons in 12~GeV $p$ + $A$ collisions}

\author{T.~Tabaru}
\email{tsugu@riken.jp}
\author{H.~En'yo}
\author{R.~Muto}
\author{M.~Naruki}
\author{S.~Yokkaichi}
\affiliation{RIKEN, 2-1 Hirosawa, Wako, Saitama 351-0198, Japan}

\author{J.~Chiba}
\altaffiliation[Present Address: ]{Department of Physics, Faculty of Science and Technology, Tokyo University of Science, 2641 Yamazaki, Noda, Chiba 378-8510, Japan}
\author{M.~Ieiri}
\author{O.~Sasaki}
\author{M.~Sekimoto}
\author{K.~H.~Tanaka}
\affiliation{Institute of Particle and Nuclear Studies, KEK, 1-1 Oho, Tsukuba, Ibaraki 305-0801, Japan}

\author{H.~Funahashi}
\author{Y.~Fukao}
\author{M.~Kitaguchi}
\author{M.~Ishino}
\altaffiliation[Present Address: ]{ICEPP, University of Tokyo, 7-3-1 Hongo, Tokyo 113-0033, Japan}
\author{H.~Kanda}
\altaffiliation[Present Address: ]{Physics Department, Graduate School of Science, Tohoku University, Sendai 980-8578, Japan}
\author{S.~Mihara}
\altaffiliation[Present Address: ]{ICEPP, University of Tokyo, 7-3-1 Hongo, Tokyo 113-0033, Japan}
\author{T.~Miyashita}
\author{K.~Miwa}
\author{T.~Murakami}
\author{T.~Nakura}
\author{F.~Sakuma}
\author{M.~Togawa}
\author{S.~Yamada}
\author{Y.~Yoshimura}
\affiliation{Department of Physics, Kyoto University, Kitashirakawa, Sakyo-ku, Kyoto 606-8502, Japan}

\author{H.~Hamagaki}
\author{K.~Ozawa}
\affiliation{Center for Nuclear Study, Graduate School of Science,\\
University of Tokyo, 7-3-1 Hongo, Tokyo 113-0033, Japan}
\author{}

\date{\today}% It is always \today, today,

\begin{abstract}
The inclusive production of $\omega$ and $\phi$ mesons
is studied in the backward region
of the interaction of 12~GeV protons with polyethylene,
carbon, and copper targets.
The mesons are measured in $e^+ e^-$ decay channels.
The production cross sections of the mesons are presented
as functions of rapidity $y$ and transverse momentum $p_T$.
The nuclear mass number dependences ($A$ dependences) are found to be
$A^{0.710 \pm 0.021(\text{stat}) \pm 0.037(\text{syst})}$ for $\omega$ mesons
and $A^{0.937 \pm 0.049(\text{stat}) \pm 0.018(\text{syst})}$ for $\phi$ mesons
in the region of $0.9 < y < 1.7$ and $p_T < 0.75$~GeV/$c$.
\end{abstract}

\pacs{13.85.Ni, 13.85.Qk, 14.40.Cs}

\maketitle

\section{\label{sec:introduction}Introduction}
The modification of the vector meson spectral function in
hot and/or dense matter is currently a hot subject
in terms of spontaneous chiral symmetry breaking
and partial restoration of the symmetry in nuclear matter.
Currently, this is the focus of several
experiments~\cite{ceres,tagx,pionic,taps}.
The experiment KEK-PS E325 was performed
to measure the vector meson spectral functions in dense matter, i.e., nucleus.
Thus far, we have reported the signature of the mass modification of
vector mesons~\cite{ozawa,naruki,muto}.
These observations are fairly remarkable; hence,
we also performed analyses to determine
the absolute cross sections and nuclear mass number dependences of
the production of these mesons
in order to understand the underlying production mechanism.

The nuclear mass number dependence of the cross sections
for the particle production is usually parameterized as
\begin{equation}
  \sigma(A)=\sigma_0 A^\alpha
  \label{eq:alpha}
\end{equation}
for a target nucleus with mass number $A$.
When the collision energy $\sqrt{s_{NN}}$ is
sufficiently large,
the parameter $\alpha$
in the production of light mesons like
pions or $\rho$ mesons is about 2/3~\cite{Ono,Binkley}.
This can be interpreted by considering such productions to be
dominated by primary collisions at the front surface of a target nucleus.
Note that the mean free path of incident protons in a nucleus is
as small as 1.4~fm.
On the other hand,
$\alpha$ tends to unity
in the case of hard reactions like the production of $J/\psi$ at high energies,
$\sqrt{s_{NN}} \gtrsim 20$~GeV~\cite{Binkley,NA50,HERA-B}.

The present experiment was performed at $\sqrt{s_{NN}} = 5.1$~GeV.
At higher energy,
$\alpha$ for $\phi$ meson production was reported
to be $0.81 \pm 0.06$, $0.96 \pm 0.04$, and $0.86 \pm 0.02$
at $\sqrt{s_{NN}} = 11.6$~\cite{Aleev},
14.2~\cite{Daum}, and 15.1~GeV~\cite{Bailey}, respectively.
However, there is no reason to believe that
these values are applicable at our energy.
A few heavy ion induced experiments
at $\sqrt{s_{NN}}$ from 4.9 to 5.4~GeV~\cite{E917,E802}
reported $\phi$ meson production data,
to which the present experiment can provide complementary data.
Note that there have been no measurements of $\omega$ mesons
and $\phi$ mesons with $p$ + $A$ reactions at $\sqrt{s_{NN}}$ around 5~GeV.

A $\phi$ meson is almost a pure $s \bar s$ state.
Therefore,
the production of a $\phi$ meson
without other accompanying strange particles
is suppressed by the Okubo-Zweig-Iizuka rule~\cite{OZI}.
This results in the effective threshold energy being as high as
$\sqrt{s_{NN}} = 3.9$~GeV, which corresponds to $2m_p + 2m_{s \bar s}$,
since two $s \bar s$ pairs are effectively needed
to realize the $\phi$ production.
Our collision energy
is fairly close to this effective threshold.
In addition, the importance of additional mechanisms for
$\phi$ meson production,
such as an intrinsic $s \bar s$ component in nucleons~\cite{Ellis}
and $\phi \rho \pi$ coupling~\cite{vonRotz,Nakayama2},
have been suggested by theorists;
however, they have thus far been insufficiently studied at our energy
from an experimental viewpoint.
Thus, to understand the production mechanism at our energy,
basic measurements such as production cross sections and $\alpha$ parameters
are indispensable.

In this article, we present inclusive production cross sections
and $\alpha$ parameters of
$\omega$ and $\phi$ mesons measured
via the $e^+ e^-$
decay channels in
12~GeV $p + p$, $p$ + C, and $p$ + Cu collisions.
The results are compared to those from nuclear cascade simulation
{\scshape jam}~\cite{jam},
and the implications to production mechanisms are discussed.

\section{\label{sec:experiment}Experiment}
The spectrometer was built in the EP1-B beam line at
the 12~GeV proton synchrotron in KEK.
The beam line was designed to deliver a 12~GeV primary proton beam
with an intensity of up to $4 \times 10^9$ protons per spill.
The beam was extracted for 1.8~s with a repetition rate of 1/4~Hz.
The beam intensity was monitored with about 10\% accuracy
using ionization chambers~\cite{IonChamber} located
downstream of the spectrometer.

\begin{figure}
  \includegraphics[scale=0.40]{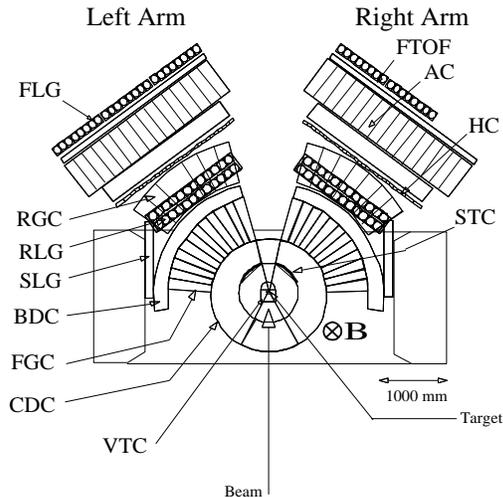}% Here is how to import EPS art
  \caption[Schematic view of experimental setup.]{ \label{fig:setup}
    Schematic view of the experimental setup from the top,
    which is designed symmetrically with respect to the beam.
    The components of this setup are
    referred as left and right arms in this article.
  }
\end{figure}

\begin{figure}
  \includegraphics[scale=0.50]{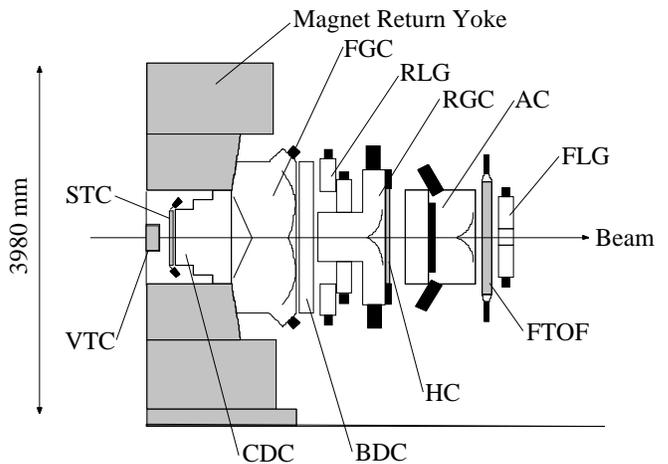}% Here is how to import EPS art
  \caption[Schematic view of experimental setup.]{ \label{fig:setup2}
    Cross section of the experimental setup
    along the plane of the center of the kaon identification counters.
  }
\end{figure}

Figures~\ref{fig:setup} and \ref{fig:setup2} show
a top view and a side view of the experimental setup, respectively.
The spectrometer was designed to simultaneously
measure $e^+ e^-$ and $K^+ K^-$ pairs.
A dipole magnet and tracking devices were commonly used
together with
electron and kaon identification counters.

The magnet was operated at 0.71~T at the center of the dipole gap
and provided 0.81~T\,m of field integral for the tracking.
The magnetic field map was calculated
by the finite-element-analysis software {\scshape tosca}~\cite{TOSCA}.
The calculated map agreed well with the measured map,
and the difference was negligible when compared to the momentum resolution
of this spectrometer.
During the data collection periods,
the field strength was monitored every 4~s
with a nuclear-magnetic-resonance (NMR) probe located at the center
of the surface of the lower pole piece.
The magnetic field map was scaled run by run according to the NMR data.
The fluctuation of the magnetic field was found to be less than
$10^{-5}$ within a typical run of two hours long.

The targets were aligned inline along the beam axis at the center of the magnet.
The target materials and configurations are shown in Table~\ref{table:config}.
\tabcolsep=6.8mm
\begin{table*}
  \begin{center}
    \caption[Target Information]{
      \label{table:config}
      Summary table for the targets, beams, and trigger modes
      used in the present analyses.
      In the 2002 run, four Cu targets were used.
    }
    \begin{tabular}{ c c d r@{.}l r@{.}l d@{ }l c }
      \hline \hline
      year & target
		& \multicolumn{1}{@{}c@{}}{\begin{tabular}{c} position \\
		(mm) \end{tabular}}
		& \multicolumn{2}{@{}c@{}}{\begin{tabular}{c} interaction \\
		length (\%) \end{tabular}}
		& \multicolumn{2}{@{}c@{}}{\begin{tabular}{c} radiation \\
		length (\%) \end{tabular}}
		& \multicolumn{2}{@{}c@{}}{\begin{tabular}{c} number \\
		of protons \end{tabular}}
		& trigger \\
      \hline
      & CH$_2$ & $-48$ & 0 & 111 & 0&195 & & & electron \\
      1999 & C & 38 & 0 & 106 & 0&213 & 3.0 & $\times 10^{13}$
      & / kaon \\
      & Cu & $-7$ & 0 & 0391 & 0&412 & & & \\
      \hline
      2002 & C & 0 & 0 & 213 & 0&431 & 3.2 & $\times 10^{14}$ & electron \\
      & Cu & \multicolumn{1}{@{}c@{}}{$\pm$24, $\pm$48} & $4 \times 0$ & 0539
		& $4 \times 0$&565 & & \\
      \hline \hline
    \end{tabular}
  \end{center}
\end{table*}

The beam profile in the horizontal direction was measured by
counting the interaction rate by changing the beam position at the target.
The beam position was moved by the bending magnet located
about 10~m upstream of the target.
In this measurement,
the center target, whose thickness was 1~mm,
was rotated by $90^\circ$ around the vertical axis,
and was used as a 1~mm-wide probe.
The typical beam size in the horizontal direction
was found to be 2.0~mm in full-width half-maximum.
The beam size in the vertical direction was known to be almost the same
as in the horizontal direction
as seen in a view of a luminescence plate
which was temporarily inserted during the beam tuning.

Three tracking devices---the vertex tracking chamber (VTC),
cylindrical drift chamber (CDC), and barrel drift chamber (BDC)---were
used to determine the trajectories of the charged particles.
In the present analysis, the momentum was determined using CDC and BDC.
The momentum resolution $\sigma_p$ was
$\sigma_p = \sqrt{(1.37\% \cdot p)^2 + 0.41\%^2}\cdot p$~(GeV/$c$),
where $p$ is the momentum of a particle.

For the electron identification,
two types of gas \v Cerenkov counters (FGC and RGC)
and three types of
lead glass calorimeters (SLG, RLG, and FLG) were employed.
The gas \v Cerenkov counters were horizontally segmented into 6$^\circ$.
The radiator of the gas \v Cerenkov counters
was isobutane at room temperature and atmospheric pressure.
The refractive index is 1.00127 at the standard temperature and pressure,
which corresponds to a momentum threshold of 2.26~GeV/$c$
for charged pions.
All lead glass detectors were built with SF6W~\cite{SF6W}
and were typically segmented to 3.5$^\circ$ horizontally.
The typical energy resolution for 1~GeV electrons is 0.15~GeV.
The acceptance for electrons
ranged from $\pm$12$^\circ$ to $\pm$90$^\circ$ in the horizontal direction
and from $-22^\circ$ to 22$^\circ$ in the vertical direction.

The counters for the kaon identification were start timing counters (STC),
hodoscopes (HC), aerogel \v Cerenkov counters (AC),
and time of flight counters (FTOF).
The acceptance for kaons ranged from $\pm$12$^\circ$ to $\pm$54$^\circ$
in the horizontal direction,
and from $-6^\circ$ to 6$^\circ$ in the vertical directions.
The three counters (STC, AC, and FTOF)
were horizontally segmented to 6$^\circ$,
and HC was typically segmented horizontally to 3$^\circ$.
The time of flight (TOF) of charged particles was measured
using STC and FTOF
with a resolution of 0.36~ns and flight length of about 3.7~m.
Kaons and pions were separated in a momentum range from 0.53 to 1.88~GeV/$c$
using an aerogel with a refractive index of 1.034.

The electron trigger signal required
a hit in FGC accompanied by a geometrical coincidence
with RGC, SLG, or RLG.
To select events containing $e^+ e^-$ pairs with large opening angles,
both the left and the right arms were required to
contain at least one $e^+$ or $e^-$ candidate.
The typical efficiency of the trigger
for the electron pairs in the acceptance
was 92.4\%.

The kaon trigger signal was obtained from the  coincidence of STC, HC, and FTOF.
The charged pion contamination was reduced by using AC as a veto trigger.
Proton contamination was reduced by setting a TOF window for the kaons
by using
a rough momentum value calculated by combining the hits in STC, HC, and FTOF
in the trigger.

The number of recorded events for electron data were
$7.41 \times 10^7$ and $5.08 \times 10^8$ in 1999 and 2002, respectively.
A detailed description of the spectrometer can be found in Ref.~\cite{nim}.

\section{\label{sec:analysis}Data Analysis}

\subsection{\label{ssec:e analysis 2002}
  Analysis of $\omega$ and $\phi \to e^+ e^-$ in $p$ + C and $p$ + Cu collisions
}
\subsubsection{\label{sssec: track reconstruction}
  Event reconstruction
}
The charged tracks were reconstructed from
hit positions in the drift chambers
by using the Runge-Kutta fitting method.
After the tracks were reconstructed,
tracks corresponding to momenta between 0.4 and 2.0~GeV/$c$ were selected
for further analyses.
The lower limit in the momentum range was set based on the threshold
of the trigger,
whereas the upper was set in order to avoid pion contamination
above the gas \v Cerenkov threshold.

\begin{figure}
  \includegraphics[scale=0.40]{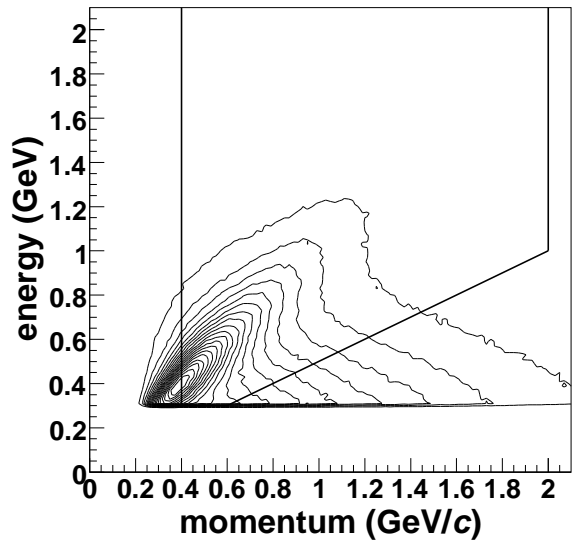}% Here is how to import EPS art
  \caption{ \label{fig:E/p}
    A contour plot of the distribution of energy and momentum of
    the present data with FGC associations.
    The solid line shows the criterion of electron identification
    described in this article.
  }
\end{figure}

Pairs of positive and negative tracks were required
to satisfy the trigger condition.
All the $e^+$ and $e^-$ candidates were reexamined so that
an FGC hit association could be obtained
with an RGC, SLG, or RLG association, depending on
the location of the track.
For candidates associated with lead glass calorimeters,
the momentum ratio $E/p$ should be larger than 0.5 for energy in
the calorimeters to be obtained.
We chose the value of 0.5 balancing the purity of $e^+$ and $e^-$
with the statistics of the present data.
Figure~\ref{fig:E/p} shows the distribution of the energy and momentum of
the present data with FGC associations.
In this figure,
it is clearly seen that electrons range along the line of $E=p$.
After an $e^+ e^-$ pair was detected,
we simultaneously refit
the $e^+$ and $e^-$ tracks by constraining them to have the same vertex point
on the interaction target.
Finally, we identified $5.69 \times 10^5$ $e^+ e^-$ pairs.\footnote{
  In a further analysis, we used only events
  in which $e^+$ went into the left arm and $e^-$ went into the right arm
  because the number of events satisfying the opposite criterion
  comprised only 6\% of the data.
}

The reconstruction efficiency of the tracks from the targets
was evaluated by both an eye-scan and
a detector simulation using {\scshape geant4}~\cite{g4}.

In the former method, first we used only one of the two arms
and determined the target in which an interaction occurred.
Then we visually scanned about 200 event displays
and found track candidates in the other arm
using the drift chamber hit information
with help from the interaction point.
The tracking efficiency was evaluated by seeing if the tracking program
could find those eye-scanned track candidates.
The efficiency was found to be 67\% on average.

In the detector simulation method,
the reconstruction efficiency was evaluated using
simulation tracks embedded onto the real data.
According to this method, the efficiency was found to be 78\%.

In order to combine the results of the two evaluations,
we simply assumed the average 73\% to be the efficiency
in the present analysis
and considered half the difference as a systematic error.
The reason of this discrepancy is unknown,
and this is one of the major sources of the systematic uncertainty
in the present data analysis.

The inefficiency of the vertex reconstruction process was evaluated as follows.
We took all the combinations of $e^+$ and $e^-$ tracks
regardless of the position of the closest point of each pair,
and obtained the yield of the $\omega$ meson peak.
Then we used the vertex reconstruction program
and the required the event vertex belong to any of the target disks.
In addition,
we fit the $e^+$ and $e^-$ tracks together
by constraining them to have the same vertex point
on the interaction target,
and required that the $\chi^2$ over the number of degrees of freedom (NDF)
should be less than 5.
As a result, we lost 6.0\% of the $\omega$ meson yield,
so that the vertex reconstruction efficiency was 94.0\%.

Contaminations due to the misidentification of pions and other particles
in the present data were evaluated at the mass region of the $\omega$ meson,
i.e., from 0.75 to 0.80~GeV/$c^2$,
by tightening the electron identifications with
gas \v Cerenkov counters and lead glass calorimeters
until the misidentification becomes negligibly small.
In this mass region, in the $p$ + C data,
we found that 18\% of the events result from
misidentification and 18\% are from uncorrelated $e^+ e^-$ pairs.

\subsubsection{\label{sssec:correction}Corrections}
Besides the tracking efficiency, several detector effects were
evaluated and corrected as described below.
The efficiencies of the electron identification counters were
evaluated as a function of momentum
by using pure electron samples from $\gamma$ conversions and Dalitz decays.
These electron samples were identified as a zero-mass peak
in the $e^+ e^-$ spectra,
and they were not required to participate in the trigger to avoid trigger bias.
The obtained efficiencies were typically 85\% for FGC,
86\% for RGC, and 97\% for the lead glass counters.

The energy losses of the tracks through the detectors
were estimated by using {\scshape geant4} simulation.
Typically, the reconstructed momentum gave a value lower by 3~MeV/$c$
for a 1~GeV/$c$ electron due to the energy loss.
The momentum difference was corrected track by track
by using a correction table obtained by the simulation.
It should be noted that this correction compensated only mean energy loss.
For effects that cause an eventual large energy loss like bremsstrahlung,
the corrections were carried out in a different manner, as described later.

\begin{figure*}
  \includegraphics[scale=0.45]{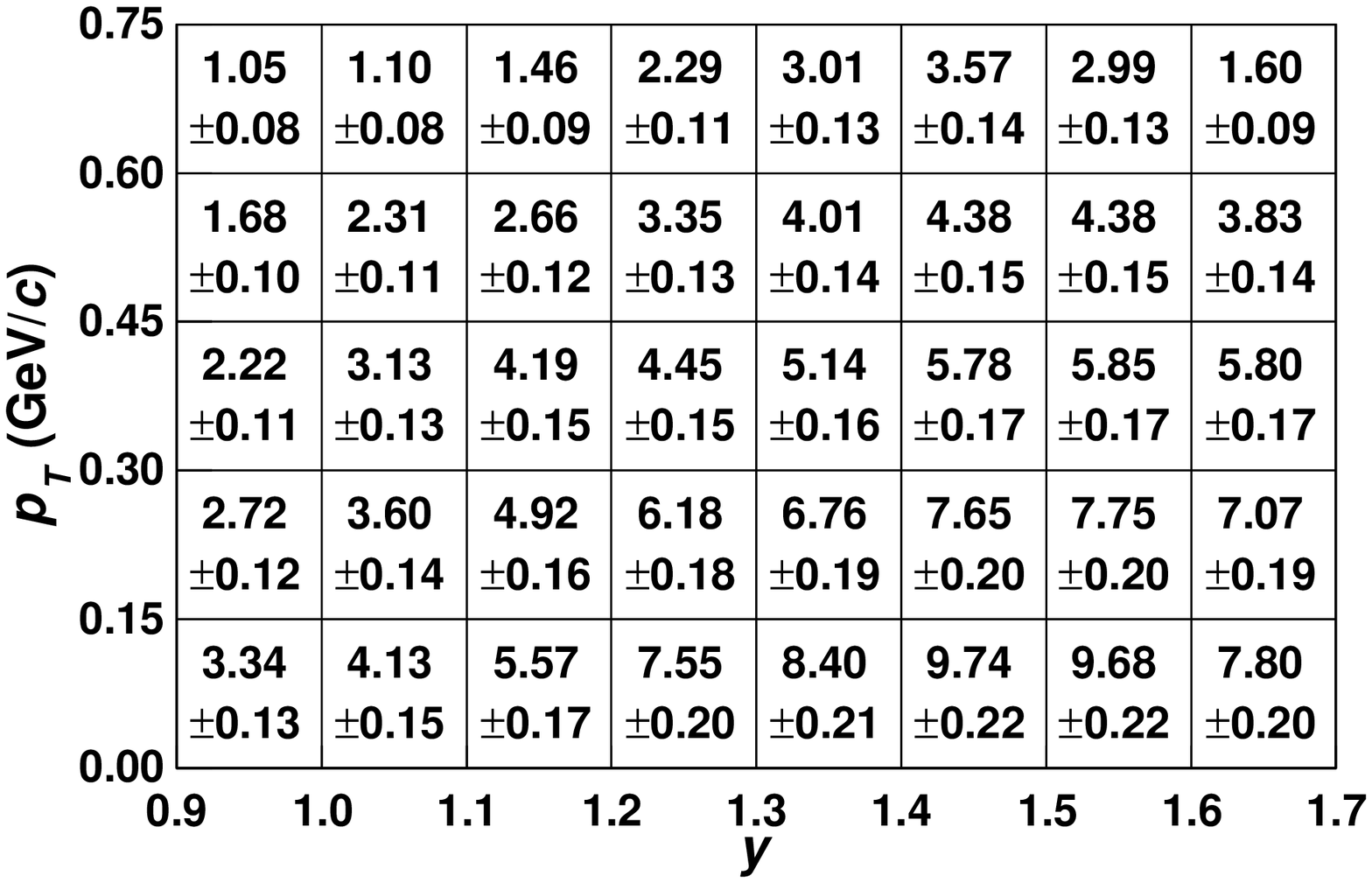}% Here is how to import EPS art
  \includegraphics[scale=0.45]{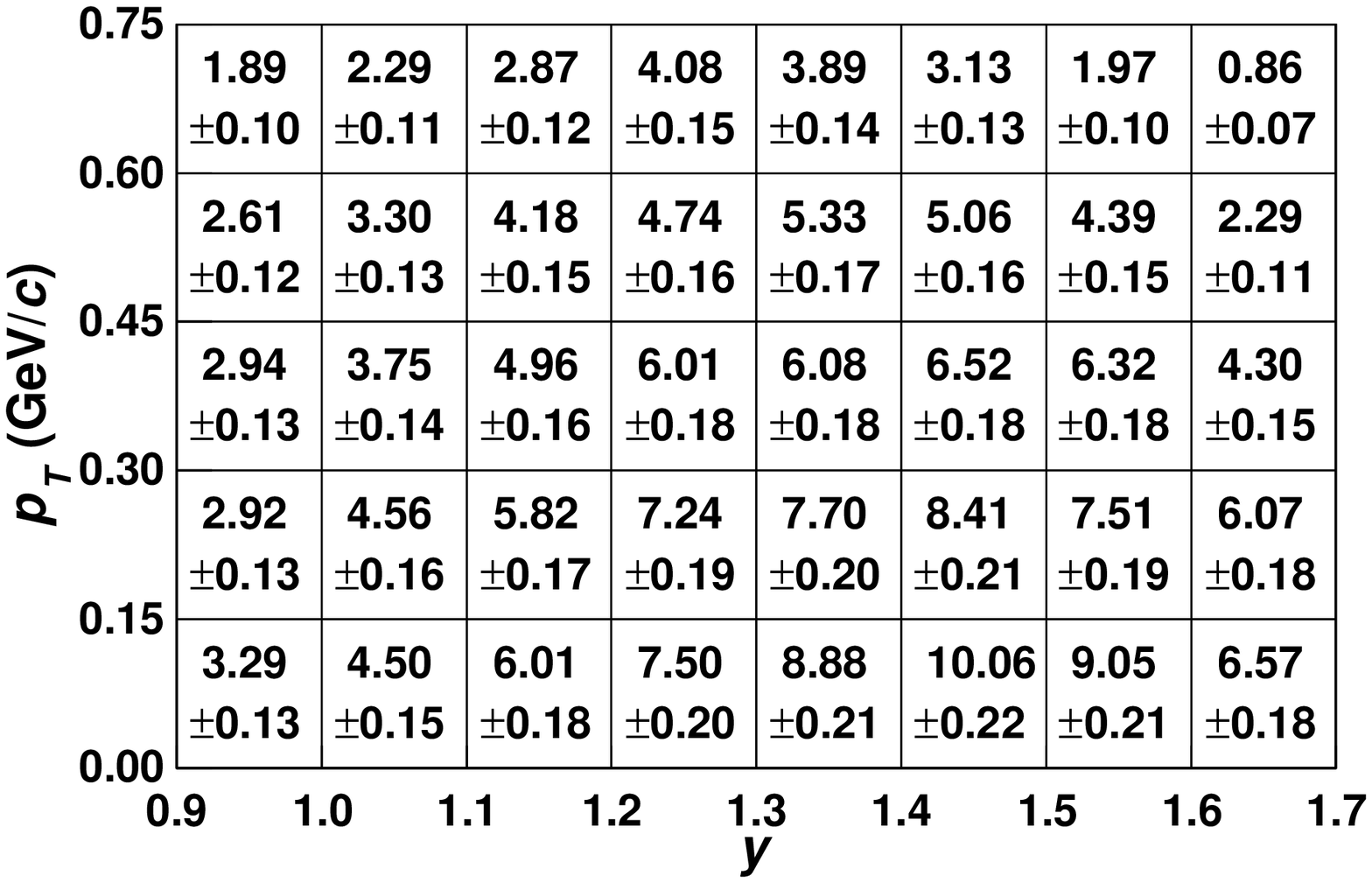}% Here is how to import EPS art
  \caption{ \label{table:Acc Sample}
    Acceptances (\%) for $\omega$ meson mass (left)
    and $\phi$ meson mass (right)
    as functions of $y$ and $p_T$ of the mesons.
    The errors are statistical and are obtained from the simulation.
  }
\end{figure*}

The geometrical acceptances for vector mesons $V$ were
obtained as functions of the invariant mass, rapidity $y$,
and transverse momentum $p_T$ by the simulation.
The acceptances were averaged over azimuth $\varphi$,
and isotropic decays of $V \to e^+ e^-$ were assumed.
The effects of the trigger, i.e., requirements of a geometrical
correlation between the electron identification counters,
were also considered.
In order to obtain the yields of vector mesons,
mass spectra were corrected for the acceptance
in the mass range above 0.55~GeV/$c^2$.
Below 0.55~GeV/$c^2$, the correction was too large to evaluate reliable values.
The obtained acceptances at the $\omega$ and the $\phi$ meson masses
are tabulated in Fig.~\ref{table:Acc Sample}.

\subsubsection{\label{sssec:spectrum decomposition}Spectrum decomposition}
\begin{figure*}
  \includegraphics[scale=0.45]{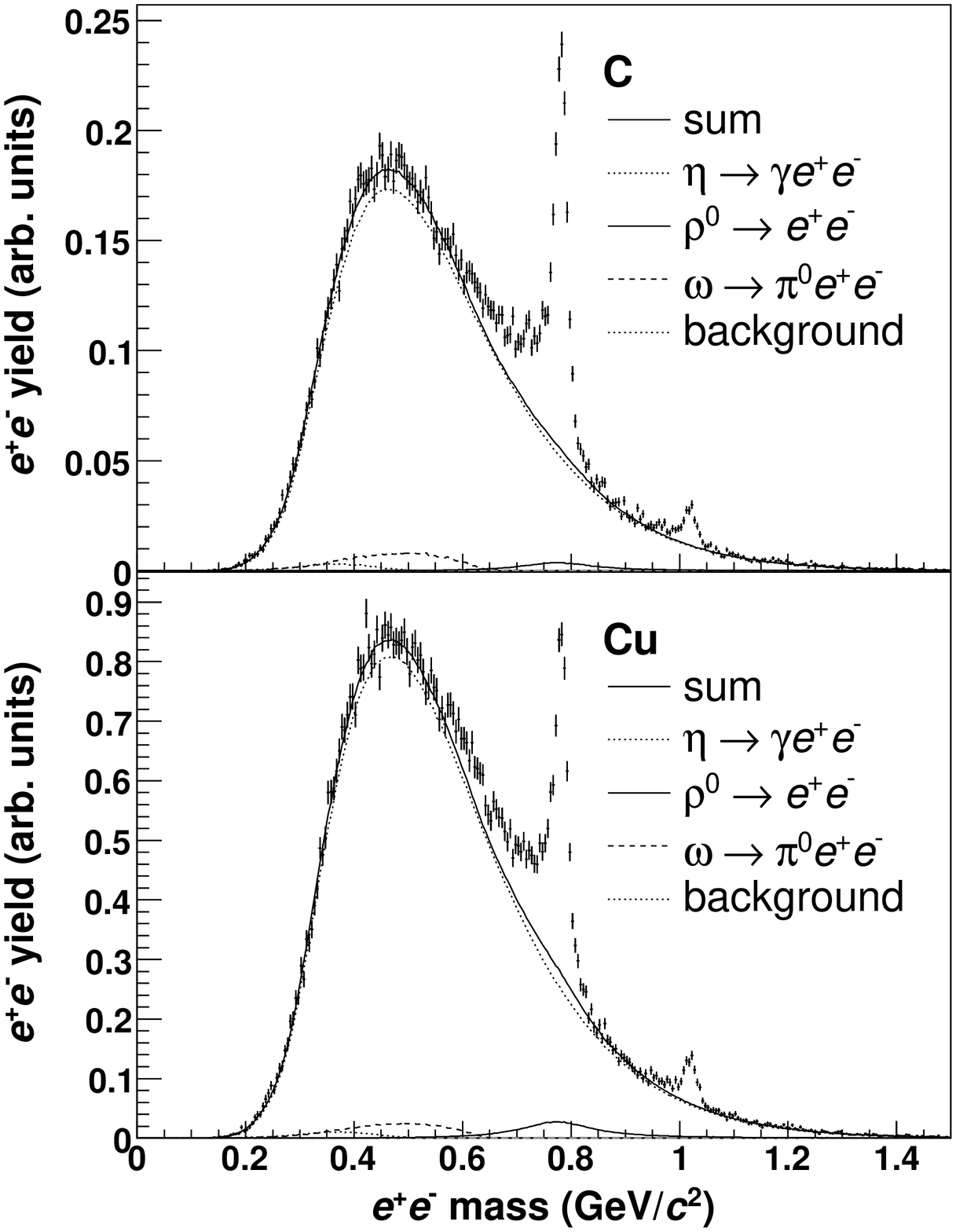}% Here is how to import EPS art
  \includegraphics[scale=0.45]{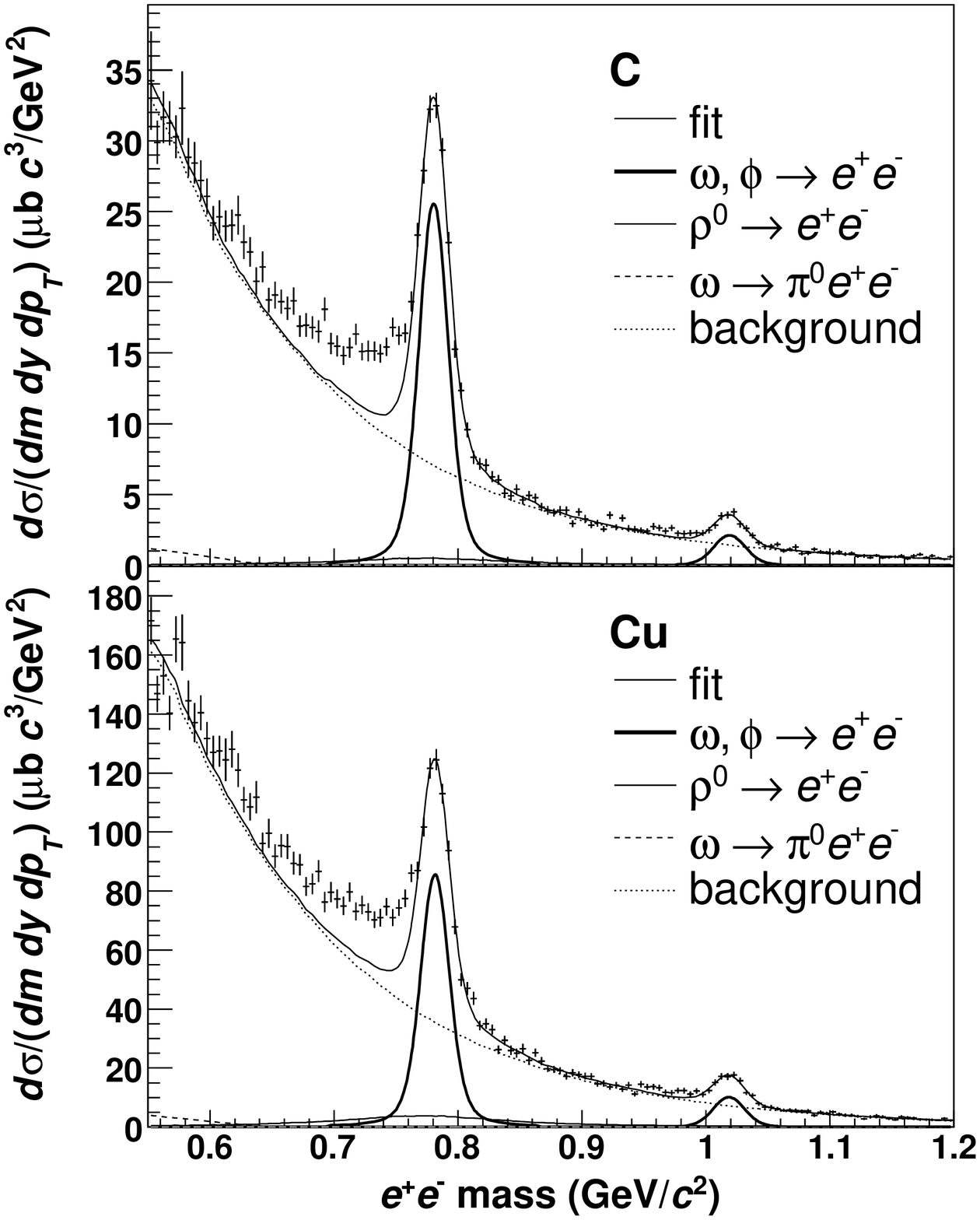}% Here is how to import EPS art
  \caption[Mass distribution before acceptance correction]{
    \label{fig:ee uncorr mass}
    The $e^+ e^-$ mass distributions of the acceptance-uncorrected data (left)
    and the corrected data (right)
    in the kinematic range of $0.9 < y < 1.7$ and $p_T < 0.75$~GeV/$c$.
    The lines in the left panels represent the backgrounds,
    Dalitz components, $\rho^0$ components, and their sum.
    The amount of $e^+ e^-$ decays of
    $\omega$ and $\phi$ mesons is represented by thick solid lines
    in the right panels,
    and the thin solid lines represent the sum of all the components.
    The number of $\phi$ meson Dalitz decays is negligibly small.
  }
\end{figure*}

In Fig.~\ref{fig:ee uncorr mass},
the left and the right panels show the spectra
of the invariant mass of $e^+ e^-$ pairs
in the range of $0.9 < y < 1.7$ and $p_T < 0.75$~GeV/$c$ without and
with the acceptance correction, respectively.
The peaks of $\omega$ and $\phi$ are distinctly observed.

The $e^+ e^-$ mass spectra were fit and decomposed
into the dielectron decays $\omega \to e^+ e^-$, $\phi \to e^+ e^-$,
and $\rho^0 \to e^+ e^-$;
the Dalitz decays $\eta \to \gamma e^+ e^-$,
$\omega \to \pi^0 e^+ e^-$,
$\phi \to \pi^0 e^+ e^-$, and $\phi \to \eta e^+ e^-$;
and the combinatorial background.
The origins of the combinatorial background were pairs
which were picked up from two independent Dalitz decays
or $\gamma$ conversions,
and pairs like $e^\pm \pi^\mp$ due to the misidentification.
The Dalitz decay $\pi^0 \to \gamma e^+ e^-$ contribution
is negligible in the acceptance of the present data.

For the invariant mass distributions of the $\omega \to e^+ e^-$
and $\phi \to e^+ e^-$ decays, we used in the fit a Breit-Wigner function
\begin{equation}
  \frac{d\sigma}{dm} = \frac{N}
		{\left( m - m_0 \right)^2 + \Gamma^2_\text{tot}/4}
  \label{eq:BW}
\end{equation}
convoluted with a Gaussian function
for the experimental resolution.
If the relativistic Breit-Wigner shape is used instead,
the results do not change significantly.
Here, $\sigma$ is the cross section; $m$, the
invariant mass of the $e^+ e^-$ pair;
$N$, a normalization factor;
$m_0$, the meson mass;
and $\Gamma_\text{tot}$, the natural decay width.
For the shape of the $\rho^0 \to e^+ e^-$ decay,
we used the relativistic Breit-Wigner shape
\begin{equation}
  \frac{d\sigma}{dm} = \frac{N}
		{\left( m^2 - m^2_{\rho^0} \right)^2 + m^2 \Gamma^2_\text{tot}},
  \label{eq:Naruki}
\end{equation}
instead of the Breit-Wigner function.
To obtain the mass distribution in the spectrometer acceptance,
we used the {\scshape geant4} simulation with
an input momentum distribution of $\rho^0$ mesons that was
obtained using {\scshape jam}.
The $e^+ e^-$ invariant mass spectra from the Dalitz decays
of $\eta$, $\omega$, and $\phi$ mesons
were also obtained by the simulation.
Their $e^+ e^-$ distributions were determined by
following the vector meson dominance model given in Ref.~\cite{VDMFormula}
by using the mother meson distributions obtained by {\scshape jam}.
The combinatorial background shape was evaluated using an event mixing method,
combining $e^+$ and $e^-$ tracks picked from different events.

The free parameters of the fit were
the yields, the peak positions, and the mass resolutions
of $\omega$ and $\phi$ mesons;
the yields of $\eta$ and $\rho^0$ mesons;
and the number of the background events.
As mentioned earlier,
the spectra were corrected in a mass range only above 0.55~GeV/$c^2$.
Therefore, below 0.55~GeV/$c^2$,
the uncorrected spectra were used in the fit
mainly to accurately obtain the amount of the background.
Although the fit region was from 0.20~GeV/$c^2$ to 1.2~GeV/$c^2$,
the mass regions from 0.600 to 0.765~GeV/$c^2$
and from 0.955 to 0.985~GeV/$c^2$ were excluded in order to avoid
the effect of the excesses below the $\omega$ and $\phi$ peaks,
which in other publications~\cite{ozawa,naruki,muto}
we have claimed as the signal of the mass modification.
In the present analysis, however,
we did not assume any underlying physics for the excess,
and we aimed only to obtain the yields of $\omega$ and $\phi$ mesons correctly,
as discussed later.

The fit results are shown by thin solid lines in Fig.~\ref{fig:ee uncorr mass}.
The $\chi^2/\text{NDF}$ were obtained as
157/152 and 192/152 for the $p$ + C and $p$ + Cu interactions, respectively.

After obtaining the raw yields of the $\omega$ and $\phi$ mesons by the fit,
the correction for
hard energy losses such as bremsstrahlung or a radiative tail was applied.
The hard energy loss causes a low mass tail
in the invariant mass distribution.
The loss of the yield due to
this tail could not be evaluated using the procedure described above;
hence, we performed studies using the {\scshape geant4} simulation.
The yields of the tails with respect to the integrals
of the Breit-Wigner peaks
were found to be $11.9 \pm 1.0$\% for the $\omega$ mesons
and $10.4 \pm 0.8$\% for the $\phi$ mesons.
These values were simply added to the peak yields.
The uncertainty of these corrections includes
an ambiguity with respect to the cross sections of
such hard energy losses in {\scshape geant4}.

By the fit procedure described above,
the peak position and mass resolution of the $\phi$ meson were
obtained as
$1019 \pm 1$~MeV/$c^2$ and $11.8 \pm 1.0$~MeV/$c^2$, respectively.
The peak position of the $\phi$ meson is consistent
with the values of the Particle Data Group~\cite{PDG},
and the mass resolution is consistent with
the simulation value of 10.7~MeV/$c^2$;
however, the peak position of the $\omega$ meson was found to be
lower by 2.2~MeV/$c^2$.
The width of the omega mesons was also broader than
that expected by the simulation.\footnote{
  If we take internal bremsstrahlung into account,
  the peak positions of both the $\omega$ mesons and the $\phi$ mesons
  agree with the expected positions.
}
Here, we adopted a conservative approach
in order to obtain the yield of unmodified
$\omega$ mesons by estimating the additional systematic error
due to this peak shift.
The error for the $\omega$ meson yield
was evaluated as 0.48\% for a $p$ + C interaction
and 1.8\% for a $p$ + Cu interaction,
which were obtained by forcing the peak position in the fit
to the higher value of 2.2~MeV/$c^2$.
The momentum scale and resolution were also verified using
the $K_s \to \pi^+ \pi^-$ data~\cite{muto}.
The measured peak position of $K_s$ was $496.8 \pm 0.3$~MeV/$c^2$,
which was consistent with that of the simulation result, 496.9~MeV/$c^2$.
The measured mass resolution of $K_s$ was $3.9 \pm 0.4$~MeV/$c^2$,
which was expected to be 3.5~MeV/$c^2$ by the simulation.

\subsubsection{\label{sssec:systematic errors}Systematic errors}
In addition to the uncertainties described previously,
the following systematic uncertainties were studied
to obtain the cross sections.

The uncertainty in the background shapes
could be a source of the systematic error.
The background was a result of uncorrelated pairs
that were obtained from two independent
Dalitz decays, $\gamma$ conversions, or other meson decays.
The background shape was obtained by an event mixing technique.
In the mixing process,
it is possible that the correlated $e^+ e^-$ pairs from the decays of $\rho^0$,
$\omega$, and $\phi$ mesons deform the estimated background shape;
this could result in a systematic error.
In order to estimate the systematic error,
two methods were used in the event mixing.
One was to use all electrons except for those belonging to
the $\omega$ and $\phi$ meson mass regions,
which are 0.765 to 0.800~GeV/$c^2$ and 0.995 to 1.035~GeV/$c^2$, respectively.
In the other method,
we used all the pairs in the event mixing but with weights
in order to obtain a self-consistent shape of the background.
The weights were obtained as a function of the $e^+ e^-$ mass
as the ratio of the background shape to the real mass spectrum,
and they were self-consistently determined
by repeating the fit several times.
We adopted the latter method to determine the background shape.
The difference between the two methods
was assigned as a systematic error.
The difference in the yield of $\omega$ mesons
was 0.23\% and that in $\phi$ mesons was 0.60\%.

The normalization and spectral shape of the combinatorial background
were affected by all the other correlated $e^+ e^-$ components
since they were obtained by the fit.
The systematic errors
in the peak-yield determination due to
the ambiguity of the Dalitz decays
of $\eta$, $\omega$, and $\phi$ mesons were evaluated
by doubling or eliminating those yields and refitting the mass spectra.
These systematic errors
were found to be 0.62\% and 0.85\% for $\omega$ and $\phi$ mesons, respectively.

It was difficult to consider the uncertainty
of the $\rho^0$ shape,
since the mass region
from 0.600 to 0.765~GeV/$c^2$ could not be represented by known sources,
and the shape of the mass modification was not well understood.
In this analysis, we simply considered the relativistic Breit-Wigner
shape for the $\rho^0$ distribution
and performed a fit by excluding the excess region.
In order to evaluate the systematic error,
we fixed the $\rho^0$ yield at zero as an extreme case
and reperformed the fit.
In this fit,
the $\omega$ yield increased by 3.19\% for the carbon target
and 7.65\% for the copper target;
further, the $\phi$ meson yield increased by 3.45\% and 5.48\%
for each of the above targets, respectively.
We considered these values as systematic errors due to the
unknown $\rho^0$ distribution.
It should be noted that the $\omega$ mesons can also be modified
such that the $\omega$ cross sections obtained in the present analysis
are only for that component whose shape is consistent with
the un-modified shape.

A possible cause of another error might lie in
the efficiency estimations for electron
identification.
These uncertainties were evaluated by using the error bands of efficiency curves
of electron identification counters
and were obtained as
2.92\% for $\omega$ mesons and 2.60\% for $\phi$ mesons.

All the systematic errors are summarized in Table~\ref{table:sys error 2002}.
In the evaluation of the absolute cross section,
the errors of the beam intensity, target thicknesses,
and efficiency of the trigger electronics were also considered.
In summary,
the systematic errors for the cross sections for $\omega$ mesons
were 19.5\% for the carbon target, and 20.8\% for the copper target,
and those of $\phi$ mesons were 19.5\% and 20.0\%, respectively.
\tabcolsep=6.6mm
\begin{table}
  \begin{center}
    \caption[Systematic Error in 2002]{
      \label{table:sys error 2002}
      Systematic uncertainties for $\omega$ and $\phi$ meson yields
      for the data taken in 2002.
    }
    \begin{tabular}{ c d@{\%} c d@{\%} }
      \hline \hline
      & \multicolumn{1}{@{}c@{}}{$\omega$}
      && \multicolumn{1}{@{}c@{}}{$\phi$} \\
      \hline
      \multicolumn{1}{@{}c@{}}{
	\begin{tabular}{c}
	  beam intensity \\
	  C target thickness \\
	  Cu target thickness \\
	  trigger electronics \\
	  track reconstruction \\
	  vertex reconstruction \\
	\end{tabular}
      } &
      \multicolumn{3}{@{}c@{}}{
	\begin{tabular}{ r@{}d@{\%} }
	  & 10 \\
	  & 0.28 \\
	  & 0.55 \\
	  & 3.8 \\
	  & 14.9 \\
	  & 4.7 \\
	\end{tabular}
      }\\
      acceptance & 1.23 && 1.25 \\
      hard energy loss & 1.18 && 0.94 \\
      electron identification & 2.92 && 2.60 \\
      background shape & 0.23 && 0.60 \\
      mass scale for $p$ + C & 0.48 && \multicolumn{1}{@{}c@{}}{} \\
      mass scale for $p$ + Cu & 1.83 && \multicolumn{1}{@{}c@{}}{} \\
      Dalitz yield & 0.62 && 0.85 \\
      $\rho^0$ from $p$ + C & 3.2 && 3.5 \\
      $\rho^0$ from $p$ + Cu & 7.6 && 5.5 \\
      \hline
      total for $p$ + C & 19.5 && 19.5 \\
      total for $p$ + Cu & 20.8 && 20.0 \\
      \hline \hline
    \end{tabular}
  \end{center}
\end{table}

The systematic errors for the $\alpha$ parameters are only due to
the items that differ between carbon and copper targets.
These are listed in Table~\ref{table:alpha sys error 2002}.
In summary, the uncertainties for the $\alpha$ parameters
of $\omega$ and $\phi$ mesons were 5.2\% and 1.9\%, respectively.
\tabcolsep=7.5mm
\begin{table}
  \begin{center}
    \caption[Systematic Error for $\alpha$ in 2002]{
      \label{table:alpha sys error 2002}
      Systematic uncertainties for $\alpha$ parameters
      of $\omega$ and $\phi$ mesons for the data taken in 2002.
    }
    \begin{tabular}{ c d@{\%} c d@{\%} }
      \hline \hline
      & \multicolumn{1}{@{}c@{}}{$\omega$}
      && \multicolumn{1}{@{}c@{}}{$\phi$} \\
      \hline
      acceptance & 1.47 && 1.13 \\
      target thickness & 0.52 && 0.39 \\
      background shape & 0.27 && 0.54 \\
      mass scale & 1.13 && \multicolumn{1}{@{}c@{}}{} \\
      Dalitz yield & 0.37 && 0.48 \\
      $\rho^0$ yield & 4.89 && 1.25 \\
      \hline
      total & 5.2 && 1.9 \\
      \hline \hline
    \end{tabular}
  \end{center}
\end{table}

\subsection{\label{ssec:e analysis 1999}
  Extraction of the cross section in $p + p$ interaction
}
The target subtraction method was employed
to obtain the cross section in $p + p$ collisions;
this was achieved by using the data taken in 1999.
The production cross sections $\sigma(p)$ in $p + p$ collisions
were calculated by the formula $\sigma(p)=(\sigma($CH$_2)-\sigma($C$))/2$
in the region of $0.9 < y < 1.7$ and $p_T < 0.6$~GeV/$c$.
The analysis was performed in almost the same manner
as that for the data in 2002.

To minimize the uncertainty due to experimental differences
between the 1999 and 2002 data,
we normalized the yield of $\omega$ mesons with carbon targets
in the 1999 data to that in 2002 data
and extracted the absolute cross sections in $p + p$ collisions.
Another uncertainty arose from the difference in the target thicknesses
that were measured with an accuracy of 0.8\%.
We obtained 20.1\% and 20.0\% as the systematic uncertainty
in $p + p$ collisions for
the $\omega$ meson production and $\phi$ meson production, respectively.

\section{\label{sec:result}Result and discussion}
\subsection{\label{ssec:raw CS}Global feature of the obtained results}
From the decay branching fractions listed in Table~\ref{table:branching ratio},
the differential cross sections of the inclusive $\omega$ meson production are
obtained as
$14.30 \pm 0.34(\text{stat}) \pm 2.79(\text{syst})$~mb$\cdot c$/GeV
and $46.63 \pm 1.25(\text{stat}) \pm 9.70(\text{syst})$~mb$\cdot c$/GeV
in $p$ + C and $p$ + Cu collisions
in the region of $0.9 < y < 1.7$ and $p_T < 0.75$~GeV/$c$, respectively;
further,
those of $\phi$ meson production are
$0.270 \pm 0.017(\text{stat}) \pm 0.053(\text{syst})$~mb$\cdot c$/GeV
and $1.290 \pm 0.070(\text{stat}) \pm 0.258(\text{syst})$~mb$\cdot c$/GeV,
respectively.\footnote{
  In this article, neither the effect of internal bremsstrahlung
  nor the uncertainty of the branching fractions is considered.
}
The $\alpha$ parameters of $\omega$ and $\phi$ mesons are obtained as
$0.710 \pm 0.021(\text{stat}) \pm 0.037(\text{syst})$
and $0.937 \pm 0.049(\text{stat}) \pm 0.018(\text{syst})$, respectively.
The difference is $0.227 \pm 0.054(\text{stat}) \pm 0.041(\text{syst})$,
and it is statistically significant.
\tabcolsep=10.0mm
\begin{table}
  \begin{center}
    \caption[Branching fractions]{ \label{table:branching ratio}
      Branching fractions in the tables of the Particle Data Group~\cite{PDG}.
    }
    \begin{tabular}{ r@{ $\to$ }l c }
      \hline \hline
      \multicolumn{2}{c}{decay mode} & branching fraction \\
      \hline
      $\omega$ & $e^+ e^-$ & $(6.96 \pm 0.15) \times 10^{-5}$ \\
      $\phi$ & $e^+ e^-$ & $(2.96 \pm 0.04) \times 10^{-4}$ \\
      $\phi$ & $K^+ K^-$ & $49.2^{+0.6}_{-0.7}$\% \\
      \hline \hline
    \end{tabular}
  \end{center}
\end{table}

In order to compare the production cross sections
in the $e^+ e^-$ decay channels with those in the $K^+ K^-$ decay channel
and the previous measurement of 12~GeV/$c$ $p + p \to \rho^0 X$
by Blobel {\it et al.}~\cite{Blobel},
the present data were
extrapolated to the backward hemisphere---region of $x_F < 0$
or $y < 1.66$---where $x_F$ is Feynman's $x$.
The production cross sections
in the backward hemisphere are listed in Table~\ref{table:CS}.
The correction factors for this extrapolation
from the measured regions---$0.9 < y < 1.7$ and $p_T < 0.75$~GeV/$c$---to
the backward hemisphere were calculated by using {\scshape jam},
since the shapes of $p_T$ and $y$ spectra are
consistent with the result of {\scshape jam} calculation, as described later.
\tabcolsep=11.4mm
\begin{table*}
  \begin{center}
    \caption[Backward cross section]{ \label{table:CS}
      Meson production cross section in the backward hemisphere.
      The first errors are statistic,
      where as the second errors are systematic.
      The data points are plotted in Fig.~\ref{fig:backwardCS}.
    }
    \begin{tabular}{ c r@{}c@{}l@{ $\pm$ }r@{}c@{}l@{ $\pm$ }r@{}c@{}l
		r@{.}l@{ $\pm$ }r@{.}l@{ $\pm$ }r@{.}l
		r@{.}l@{ $\pm$ }r@{.}l@{ $\pm$ }r@{.}l }
      \hline \hline
      & \multicolumn{9}{c}{$\omega$ (mb)}
      & \multicolumn{6}{c}{$\phi$ ($e^+ e^-$) (mb)}
      & \multicolumn{6}{c}{$\phi$ ($K^+ K^-$) (mb)} \\
      \hline
      H &  2&.&3 &1&.&1 &  0&.&5 & 0&034 & 0&045 & 0&007 & 0&03 & 0&19 & 0&01 \\
      C & 13&.&4 &0&.&3 &  2&.&6 & 0&240 & 0&015 & 0&047 & 0&39 & 0&05 & 0&19 \\
      Cu& 49&.&0 &1&.&3 & 10&.&2 & 1&21  & 0&07  & 0&24  & 1&84 & 0&27 & 0&89 \\
      Pb&\multicolumn{9}{c}{} & \multicolumn{6}{c}{} & 6&0 & 1&8 & 2&9  \\
      \hline \hline
    \end{tabular}
  \end{center}
\end{table*}

The cross sections of the inclusive $\phi$ meson production
in the $K^+ K^-$ decay channel are also listed in Table~\ref{table:CS}.
These were obtained from the previous analysis in
this experiment~\cite{yokkaichi,ishino}.
The $\alpha$ parameter of the $\phi$ meson
production measured in the $K^+K^-$ decay channel is obtained
as $1.01 \pm 0.09$ using
the $p + p$, $p$ + C, $p$ + Cu, and $p$ + Pb data
in the spectrometer acceptance;
this value is statistically consistent with the present $e^+ e^-$ analysis.

\begin{figure}
  \includegraphics[scale=0.45]{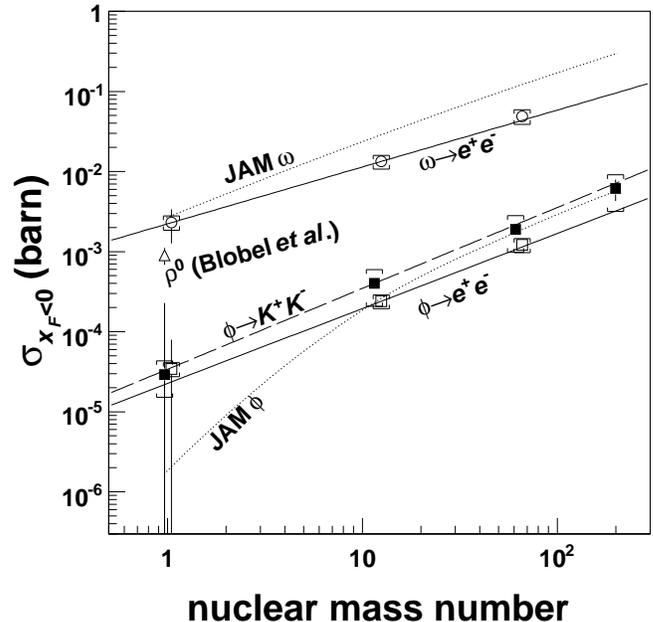}% Here is how to import EPS art
  \caption[Backward cross section]{ \label{fig:backwardCS}
    Production cross section in the backward hemisphere as a function of
    mass number.
    The circles, open squares, and filled squares show the
    cross section of $\omega$ mesons,
    $\phi$ mesons measured in the $e^+ e^-$ decay channel, and
    those in the $K^+ K^-$ decay channel,
    respectively.
    The vertical lines represent the statistical errors,
    and the brackets represent the systematic errors.
    The previous $p + p \to \rho^0 X$ measurement~\cite{Blobel}
    is indicated by a triangle.
    The dotted lines represent the prediction
    by {\scshape jam} simulation results.
    The solid and dashed lines represent the $\alpha$ parameterization
    (see text).
  }
\end{figure}

Figure~\ref{fig:backwardCS} shows the cross sections
in the backward hemisphere as a function of the nuclear mass number.
The dotted lines represent the result of the {\scshape jam} calculation
and the solid and dashed lines represent the measured $\alpha$ parameterization.
The $\alpha$ parameters shown in the figure were 0.710, 0.937, and 1.01
for the data of $\omega \to e^+ e^-$, $\phi \to e^+ e^-$,
and $\phi \to K^+ K^-$, respectively.
It should be noted that
these values were obtained from the data within the spectrometer acceptance
before they were extrapolated to the backward hemisphere.

The previous measurement~\cite{Blobel} yielded the total cross section
of $1.8 \pm 0.25$~mb for $\rho^0$ mesons.
By assuming a $\rho^0/\omega$ ratio of $1.0 \pm 0.2$,\footnote{
  The $\rho^0/\omega$ ratio was measured by
  the reactions of
  $p + p \to \rho^0$ + charged particles
  and
  $p + p \to \omega$ + charged particles~\cite{Blobel}.
}
we obtain the $\omega$ meson production in the backward hemisphere
as $0.90 \pm 0.22$~mb for a comparison with the present measurement.
The triangle in Fig.~\ref{fig:backwardCS} represents the obtained value.
The measured $\omega$ meson production cross section
in the present $p + p$ collision data
is consistent with their $\rho^0$ cross section within the error.

\subsection{\label{ssec:diff CS}
  Differential cross section of $\omega$ and $\phi$ production measured
  in $e^+ e^-$ decays
}
\begin{figure*}
  \includegraphics[scale=0.70]{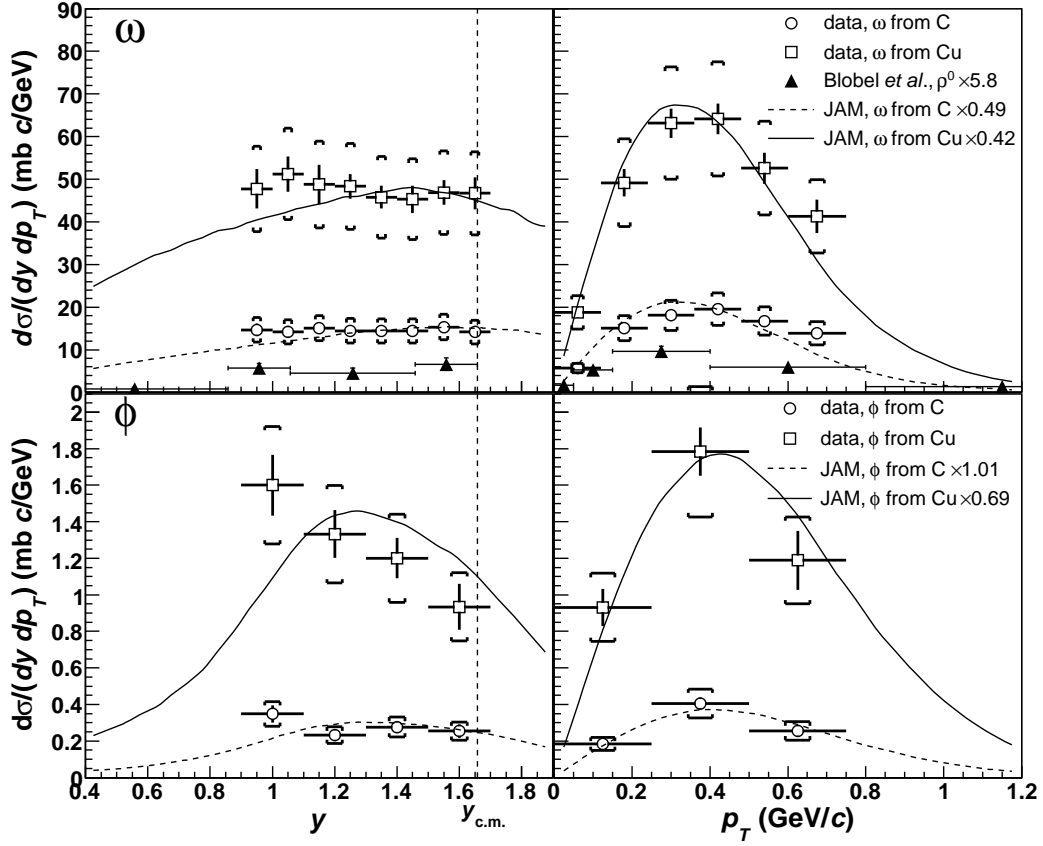}% Here is how to import EPS art
  \caption{ \label{fig:DCS}
    Differential cross sections of $\omega$ (top) and $\phi$ (bottom) mesons
    as functions of $y$ (left) or $p_T$ (right).
    The statistical errors are represented by vertical bars,
    and the systematic errors are represented by brackets.
    The previous $p + p \to \rho^0 X$ measurement~\cite{Blobel}
    is indicated by triangles
    after scaling up with the nuclear mass number dependence
    obtained in the present analysis.
    The dashed and solid curves
    represent scaled cross sections of {\scshape jam}
    in $p$ + C and $p$ + Cu collisions, respectively.
    The rapidity of the center of mass system is 1.66, as indicated by
    $y_{\text{c.m.}}$.
    The data points are listed in Table~\ref{table:DiffCS}.
  }
\end{figure*}

In a region $0.9 < y < 1.7$ and $p_T < 0.75$~GeV/$c$,
the differential cross sections
of $\omega$ and $\phi$ mesons were obtained
for each $y$ or $p_T$ bin, as shown in Fig.~\ref{fig:DCS}.
These are also listed in Table~\ref{table:DiffCS}.
\tabcolsep=2.8mm
\begin{table*}
  \begin{center}
    \caption[Differentail cross section]{ \label{table:DiffCS}
      Differential production cross sections in $0.9 < y < 1.7$
      and $p_T < 0.75$~GeV/$c$ measured via $e^+ e^-$ decay channel.
      The first errors are statistical and the second are systematic.
      The data points are plotted in Figs.~\ref{fig:DCS} and \ref{fig:alpha}.
    }
    \begin{tabular}{ c r@{.}l@{--}r@{.}l d@{ $\pm$ }r@{.}l@{ $\pm$ }r@{.}l
		d@{ $\pm$ }r@{.}l@{ $\pm$ }r@{.}l
		d@{ $\pm$ }r@{.}l@{ $\pm$ }r@{.}l }
      \hline \hline
	\multicolumn{5}{c}{}
	& \multicolumn{5}{@{}c@{}}{\begin{tabular}{c} $\omega$ from C \\
		(mb $c$/GeV) \end{tabular}}
	& \multicolumn{5}{@{}c@{}}{\begin{tabular}{c} $\omega$ from Cu \\
		(mb $c$/GeV) \end{tabular}}
	& \multicolumn{5}{c}{$\alpha$} \\
      \hline
	& 0&9 & 1&0 & 14.7 & 1&5 & 2&9 & 47.8 & 4&6 & 10&0 & 0.708 & 0&083 & 0&042 \\
	& 1&0 & 1&1 & 14.2 & 1&2 & 2&8 & 51.2 & 4&2 & 10&7 & 0.770 & 0&070 & 0&041 \\
	& 1&1 & 1&2 & 15.0 & 1&1 & 3&0 & 48.8 & 4&6 & 10&2 & 0.706 & 0&071 & 0&040 \\
	$y$
	& 1&2 & 1&3 & 14.5 & 0&8 & 2&8 & 48.3 & 2&9 & 10&1 & 0.724 & 0&049 & 0&039 \\
	& 1&3 & 1&4 & 14.4 & 0&8 & 2&8 & 45.8 & 2&7 &  9&6 & 0.693 & 0&048 & 0&039 \\
	& 1&4 & 1&5 & 14.4 & 0&7 & 2&8 & 45.3 & 3&2 &  9&5 & 0.689 & 0&052 & 0&039 \\
	& 1&5 & 1&6 & 15.3 & 0&7 & 3&0 & 46.9 & 2&9 &  9&8 & 0.671 & 0&045 & 0&038 \\
	& 1&6 & 1&7 & 14.2 & 0&7 & 2&8 & 46.7 & 3&8 &  9&7 & 0.716 & 0&056 & 0&039 \\
      \hline
	& 0&00 & 0&12 &  5.7 & 0&4 & 1&1 & 18.8 & 1&2 &  3&9 & 0.714 & 0&057 & 0&038 \\
	& 0&12 & 0&24 & 15.1 & 0&7 & 3&0 & 49.2 & 3&2 & 10&2 & 0.709 & 0&047 & 0&038 \\
	$p_T$ & 0&24 & 0&36 & 18.1 & 0&8 & 3&5 & 63.1 & 3&4 & 13&2 & 0.750 & 0&042 & 0&039 \\
	(GeV/$c$) & 0&36 & 0&48 & 19.5 & 0&9 & 3&8 & 64.2 & 3&6 & 13&4 & 0.714 & 0&044 & 0&038 \\
	& 0&48 & 0&60 & 16.8 & 1&3 & 3&3 & 52.6 & 3&6 & 11&0 & 0.687 & 0&063 & 0&040 \\
	& 0&60 & 0&75 & 13.9 & 1&1 & 2&7 & 41.3 & 3&9 &  8&6 & 0.655 & 0&075 & 0&041 \\
      \hline
	\multicolumn{5}{c}{$0.9 < y < 1.7$, $p_T < 0.75$~GeV/$c$} & 14.30 & 0&34 & 2&79 & 46.63 & 1&25 & 9&70 & 0.710 & 0&021 & 0&037 \\
      \hline
      \hline
	\multicolumn{5}{c}{}
	& \multicolumn{5}{c}{$\phi$ from C}
	& \multicolumn{5}{c}{$\phi$ from Cu}
	& \multicolumn{5}{c}{$\alpha$} \\
      \hline
	& 0&9 & 1&1 & 0.348 & 0&046 & 0&068 & 1.60 & 0&17 & 0&32 & 0.916 & 0&101 & 0&022 \\
	$y$
	& 1&1 & 1&3 & 0.232 & 0&032 & 0&045 & 1.33 & 0&13 & 0&27 & 1.050 & 0&101 & 0&020 \\
	& 1&3 & 1&5 & 0.277 & 0&029 & 0&054 & 1.20 & 0&11 & 0&24 & 0.881 & 0&084 & 0&020 \\
	& 1&5 & 1&7 & 0.255 & 0&037 & 0&050 & 0.93 & 0&13 & 0&19 & 0.780 & 0&119 & 0&019 \\
      \hline
	$p_T$
	& 0&00 & 0&25 & 0.185 & 0&024 & 0&036 & 0.93 & 0&10 & 0&19 & 0.971 & 0&101 & 0&019 \\
	(GeV/$c$)
	& 0&25 & 0&50 & 0.405 & 0&032 & 0&079 & 1.78 & 0&13 & 0&36 & 0.890 & 0&066 & 0&019 \\
	& 0&50 & 0&75 & 0.255 & 0&032 & 0&050 & 1.19 & 0&16 & 0&24 & 0.924 & 0&111 & 0&021 \\
      \hline
	\multicolumn{5}{c}{$0.9 < y < 1.7$, $p_T < 0.75$~GeV/$c$} & 0.270 & 0&017 & 0&053 & 1.290 & 0&070 & 0&258 & 0.937 & 0&049 & 0&018 \\
      \hline \hline
    \end{tabular}
  \end{center}
\end{table*}

The previous $p + p \to \rho^0 X$ measurement~\cite{Blobel}
was plotted using triangles, as shown in Fig.~\ref{fig:DCS}.
It should be noted that
the data points were scaled by factors of 5.81 (= $12^{0.71}$) for clarity.

The distributions obtained by the {\scshape jam} calculation are compared
with the measurements in Fig.~\ref{fig:DCS}.
The total cross sections obtained from the
{\scshape jam} calculation are larger than that obtained from the data.
Hence, they are scaled by the factors 0.489 for $p$ + C $\to \omega X$,
0.421 for $p$ + Cu $\to \omega X$,
1.006 for $p$ + C $\to \phi X$, and 0.686 for $p$ + Cu $\to \phi X$.
These scale factors were determined as the ratio of the total cross sections
in the acceptance between the data and {\scshape jam}.
The shapes of the differential cross sections of {\scshape jam}
are consistent with the present data,
although the absolute cross sections are
systematically larger than the data.

\begin{figure*}
  \includegraphics[scale=0.70]{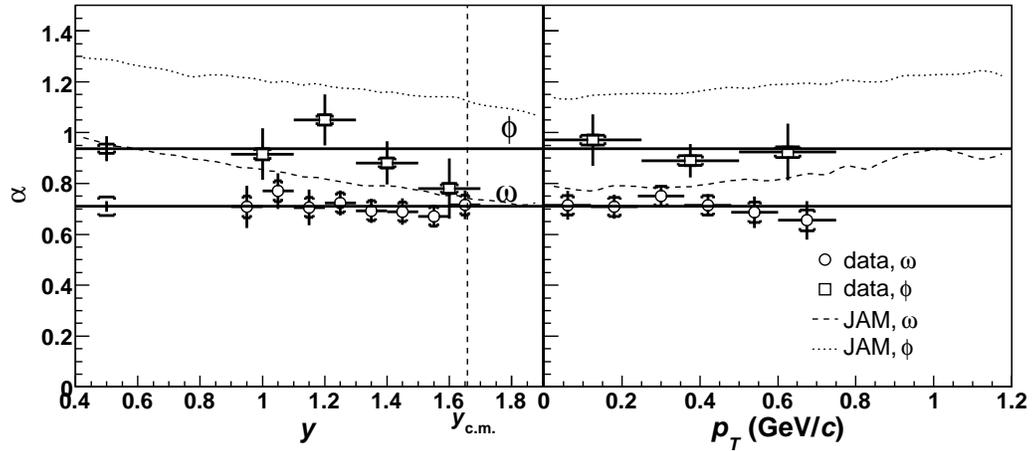}% Here is how to import EPS art
  \caption{ \label{fig:alpha}
    The $\alpha$ parameters of $\omega$ (circle) and $\phi$ (square) mesons
    as functions of $y$ (left) or $p_T$ (right).
    The statistical errors are represented by vertical bars
    and the systematic errors are represented by brackets.
    The horizontal lines
    represent the averaged $\alpha$ parameters of $\omega$ and $\phi$ mesons
    with the errors indicated on the left side.
    The dotted and dashed curves represent the
    $\alpha$ parameters of $\phi$ and $\omega$ mesons by {\scshape jam}.
    The rapidity of the center of mass system is indicated by $y_{\text{c.m.}}$.
    The data points are listed in Table~\ref{table:DiffCS}.
  }
\end{figure*}

Figure~\ref{fig:alpha} shows the $\alpha$ parameters
of $\omega$ and $\phi$ mesons obtained
in the region of $0.9 < y <1.7$ and $p_T < 0.75$~GeV/$c$.
These are also listed in Table~\ref{table:DiffCS}.
The flat lines represent the averaged $\alpha$ parameters
with the errors shown in the left column.
The dotted and dashed curves represent
the results of {\scshape jam} for $p$ + C and $p$ + Cu collisions.
Although the $\alpha$ parameters of the {\scshape jam} calculation are
significantly larger than those of the data,
the difference between $\omega$ and $\phi$ mesons in the $\alpha$ parameters
is similar to what is seen in the {\scshape jam} calculation.

\subsection{\label{ssec:discussion} Discussion}
An interesting characteristic is the difference in the $y$ dependence
between $\omega$ and $\phi$ meson production cross sections.
While the $\omega$ production is essentially independent of $y$,
the $\phi$ production increases towards the smaller $y$ values.
In addition,
the $\alpha$ parameters of $\omega$ and $\phi$ meson production
are different by $0.227 \pm 0.054(\text{stat}) \pm 0.041(\text{syst})$.
This significant difference in the $\alpha$ parameters
confirms that the production mechanisms of $\omega$ and $\phi$ mesons
are different.

One possible mechanism of $\phi$ meson production
is a hard reaction between incident and projectile nucleons.
If the hard reaction is dominant like $J/\psi$ production at higher energies,
$\alpha$ is expected to be around unity and independent of $y$.
Further, the $y$ distribution of the cross section is expected to be symmetric
with respect to $y_{\text{c.m.}}$,
which is not the case in the present $\phi$ production data.

Another possible effect to explain the observed characteristics of
the $\alpha$ parameter and cross sections
is the effect of secondary collisions in a target nucleus.
In this case, no hard reaction is necessary.
These effects are expected to increase in a smaller rapidity region
for $p$ + $A$ interaction,
and the cross sections and $\alpha$ are expected to
be larger in the backward region.

Although the {\scshape jam} does not reproduce the data quantitatively,
it yields fairly similar shapes
in the inclusive cross sections
of $\omega$ and $\phi$ meson productions;
further, it also predicts
the difference between the $\alpha$ parameters of these mesons.
In the {\scshape jam} calculation,
more than 90\% of the $\phi$ mesons are produced by secondary
collisions mostly between non-strange mesons
and nucleons in a target nucleus,
while $\omega$ mesons are produced both in primary and secondary reactions.
For example, 30\% and 50\% of the $\omega$ mesons are produced
in secondary reactions in the cases of 12~GeV $p$ + C and $p$ + Cu reactions
in {\scshape jam}, respectively.
These differences in the production mechanism
in {\scshape jam} will surely be an important step to understand
the measured differential cross sections.

The difference between the scaling factors applied to
$\phi$ meson production in $p$ + C and $p$ + Cu collisions,
in {\scshape jam}, described
in Sec.~\ref{ssec:diff CS}, can be understood qualitatively
by the overestimated contribution of the secondary collisions
in {\scshape jam}.
Although in those calculations almost all the $\phi$ mesons are produced
in secondary collisions,
the present data suggest that the contribution of the primary collisions
can be larger.
For the case of the $\omega$ meson,
{\scshape jam} just predicts larger absolute production yields than
seen in the present data.

In summary, we measured the inclusive differential cross sections
of the $\omega$ and $\phi$ meson production
in $p$ + $A$ collisions in the backward region.
The difference in the $\alpha$ parameters between $\omega$ and $\phi$ mesons
confirms that the production mechanisms of $\omega$ and $\phi$ mesons
are different.
The results are compared to the nuclear cascade calculations.

\begin{acknowledgments}
We have greatly appreciated all of the staff members of KEK-PS,
especially the beam channel group, for their helpful support.
We would like to thank C. Lourenco for proof reading of the manuscript.
This work was partly funded by the Japan Society for the Promotion of Science,
RIKEN Special Post doctoral Researchers Program,
and a Grant-in-Aid for Scientific Research from the Japan Ministry of Education,
Culture, Sports, Science and Technology (MEXT).
Finally, we also thank the staff members of RIKEN super combined cluster system
and RIKEN-CCJ.
\end{acknowledgments}

\bibliography{draft09}% Produces the bibliography via BibTeX.

\end{document}